\documentclass[%
 reprint,
 amsmath,amssymb,https://www.overleaf.com/project/62490235f741f61e5ad70f44
 aps,
]{revtex4-2}

\usepackage{graphicx}
\usepackage{dcolumn}
\usepackage{bm}
\usepackage{color}
\usepackage{url}

\newcommand \MZ [1] {\bgroup\noindent[\textcolor{blue}{\textbf{MZ}: #1}]\egroup\ignorespacesafterend}
\newcommand \SH [1] {\bgroup\noindent[\textcolor{green}{\textbf{SH}: #1}]\egroup\ignorespacesafterend}
\newcommand 
\SZ [1] {\bgroup\noindent[\textcolor{red}{\textbf{SZ}: #1}]\egroup\ignorespacesafterend}
\newcommand 
\PM [1]
{\bgroup\noindent[\textcolor{yellow}{\textbf{PM}: #1}]\egroup\ignorespacesafterend}

\begin{document}

\preprint{APS/123-QED}

\title{Fiber bundle model of thermally activated creep failure}

\author{Stefan Hiemer}
 \email{stefan.hiemer@unimi.it}
\affiliation{%
 Center for Complexity and Biosystems, Department of Physics "Aldo Pontremoli", University of Milan, Via Celoria 16, 20133 Milano, Italy
}%
\affiliation{%
CNR -- Consiglio Nazionale delle Ricerche, Istituto di Chimica della Materia Condensata e di Tecnologie per l'Energia, Via R. Cozzi 53, 20125 Milano, Italy}%
\author{Paolo Moretti}%
 \email{paolo.moretti@fau.de}
\affiliation{%
 Friedrich-Alexander-Universität Erlangen-Nürnberg, Institute of Materials Simulation, Department of Materials Science and Engineering, Fürth, Germany
}%
\author{Stefano Zapperi}%
 \email{stefano.zapperi@unimi.it}
\affiliation{%
 Center for Complexity and Biosystems, Department of Physics "Aldo Pontremoli", University of Milan, Via Celoria 16, 20133 Milano, Italy
}%
\affiliation{%
CNR -- Consiglio Nazionale delle Ricerche, Istituto di Chimica della Materia Condensata e di Tecnologie per l'Energia, Via R. Cozzi 53, 20125 Milano, Italy}
\author{Michael Zaiser}%
 \email{michael.zaiser@fau.de}
\affiliation{%
 Friedrich-Alexander-Universität Erlangen-Nürnberg, Institute of Materials Simulation, Department of Materials Science and Engineering, Fürth, Germany
}%

\date{\today}

\begin{abstract}
An equal load sharing fiber bundle model for thermally activated breakdown is developed using transition state theory to describe the rate of elementary failures. The lifetime distribution, average, variance and their asymptotic limits for uniform fiber failure thresholds are derived and found to be in excellent agreement with simulations. The asymptotic scaling with regards to the number of fibers matches analytical approximations in the low temperature limit derived by Roux and co-workers for a model of thermal breakdown by stationary Gaussian noise. For the case of randomly distributed fiber failure strengths, the lifetime distribution is derived as a multidimensional integral with no closed form solution. Simulations with different fiber strength distributions indicate that, in the limit of large fiber numbers, the statistics of bundle lifetimes shows a similar asymptotic scaling for distributed and for uniform thresholds. Fiber breakage by thermal activation occurs in avalanches triggered by individual thermally activated failure events, and the asymptotic avalanche size distribution obtained from the simulations matches earlier theoretical results derived for quasistatic loading.
\end{abstract}

\maketitle


\section{Introduction}
Understanding material failure under mechanical loading is a central challenge in science and engineering, presenting many open problems despite centuries of research dating back to Leonardo da Vinci and Galileo Galilei \cite{uccelli1940libri,galilei1958due}. Among the simplest models for capturing the statistical aspects of failure is the fiber bundle model, which remains analytically tractable and captures key features of real-world fracture. Originally invented by Peirce in the context of textile manufacturing \cite{peirce}, the fiber bundle model was expanded by Daniels into a general framework for analytical modelling of statistical fracture \cite{daniels}. In its original design, fibers are connected to two rigid bars and subjected to tensile loading with the same force acting on each fiber.  The mechanical behavior of the fibers is that of a Hookean spring typically with unit spring constant which breaks once the force exceeds the individual failure threshold of the fiber. In this simple model where load is shared equally among all the fibers, asymptotic results have been derived, revealing power-law statistics of fiber failure avalanches with an exponential cutoff depending on the threshold distribution
and the external load \cite{hemmer,sornette,kloster}. The fiber bundle model has also been extended to incorporate additional physical phenomena. To account for stress concentration effects, as observed in continuous media, various load redistribution schemes—such as the local load sharing model—have been introduced \cite{harlow,phoenix1983comparison}. Moreover, alternative constitutive laws describing the mechanical behavior of individual fibers have been analyzed (see \cite{alava,hansen} for detailed reviews of fiber bundle models). \\
The present work focuses on the special case of thermally activated breakdown. Guarino, Scoretti and Ciliberto (GSC) proposed a model based on thermal Gaussian stationary noise acting on each fiber as an additional force fluctuation \cite{guarino}. The model was later analyzed by Roux who derived asymptotic results for many identical fibers with unit threshold in the limit of low temperatures \cite{roux}. Politi and Scorretti further studied perturbative expansions for the disordered case \cite{politi,scorretti}. So far, no study derived the complete failure time distribution, which would provide a complete description of the failure
process. 

One should also note that in real materials, creep failure is typically associated with underlying deformation mechanisms characteristic of the respective material class. In metals, this may be dislocation glide, vacancy diffusion or other metal processes which are associated with an Arrhenius-type temperature dependency of the creep rate \cite{ilschner,gottstein}. In glasses, the underlying creep mechanism can be related to shear transformations  \cite{castellanosdiss}. Based on these observations, we formulate a fiber bundle model based on transition state theory \cite{eyring}, assuming an Arrhenius-type temperature dependence of the failure rate of individual fibers. We find that this assumption is consistent with phenomenological observations and is also easier to verify experimentally than the existence of a thermal Gaussian noise acting on a macroscopic object. Under this assumption, we derive the exact lifetime distribution for a system of fibers with uniform failure thresholds, as well as the asymptotic behavior of this distribution in the limit of large fiber numbers and large forces. For the case when failure thresholds are statistically distributed, we provide a solution in the form of a high-dimensional integral with (yet) unknown closed-form solution. 

\section{The model}
We assume equal load sharing where a bundle consisting of $N$ fibers is subject to a constant load $F$, so that the initial load  \(f_{0}\) per fiber is 
\begin{equation}
    f_{0}= \frac{F}{N}.
\end{equation} 
For a fiber $k$ there exist two failure modes: i) The fiber's failure threshold $t_k$ is below the force $f$ acting on the fiber,  which leads to immediate failure; ii) the fibers fails under a subcritical load $f < t_k$ after some randomly distributed time due to a thermally activated barrier crossing event. If other fibers fail immediately after such an event due to overloading, we refer to this as an avalanche and define the avalanche size as the number of fibers that fail collectively by overloading before a new metastable configuration is reached where $f < t_j \forall j$. The force per fiber for the \(i\)-th thermally activated event is  
\begin{equation}
    f = \frac{F}{n_{i}}.
\end{equation} 
where \(n_{i}\) is the number of intact fibers before the thermal event. The failure rate of an individual fiber \(k\) is taken from transition state theory \cite{eyring}. Using a linearization of the energy barrier with regards to the force, similar to the assumption made in previous finite element studies of thermally activated creep failure \cite{castellanos}, we find
\begin{equation}
    \nu_{k} = \exp\left(-\frac{t_{k}-f}{T}\right)
\end{equation}
where \(T\) is the appropriately scaled temperature. As the individual thermal events are assumed to be statistically independent with regards to time and space, the total event/avalanche rate before event $i$ is given by 
\begin{equation}
    \nu_{i,\rm tot} = \sum_{k=1}^{n_i} \exp\left(-\frac{t_{k}+f}{T}\right)
\end{equation}
and the probability of fiber \(q\) to fail via the $i$th thermal event is
\begin{equation}
    P\left(q,i\right)=\frac{\nu_{i,{\rm q}}}{\nu_{i, {\rm tot}}}
    \label{prob_chosen}
\end{equation}
The distribution of times between events is exponential 
\begin{equation}
    p(\tau_{i})= \exp\left(-\nu_{i,{\rm tot}}\tau_{i}\right),
\end{equation}
and the total failure time is the sum of the individual inter-event times
\begin{equation}
    \tau= \sum_{i=1}^{N_{\rm ev}} \tau_{i}
\end{equation}
where \(N_{\rm ev}\) is the total number of thermal events until all fibers are broken.

\section{Fibers with uniform thresholds}
\begin{figure*}
\includegraphics[width=\textwidth,height=0.35\textheight,keepaspectratio]{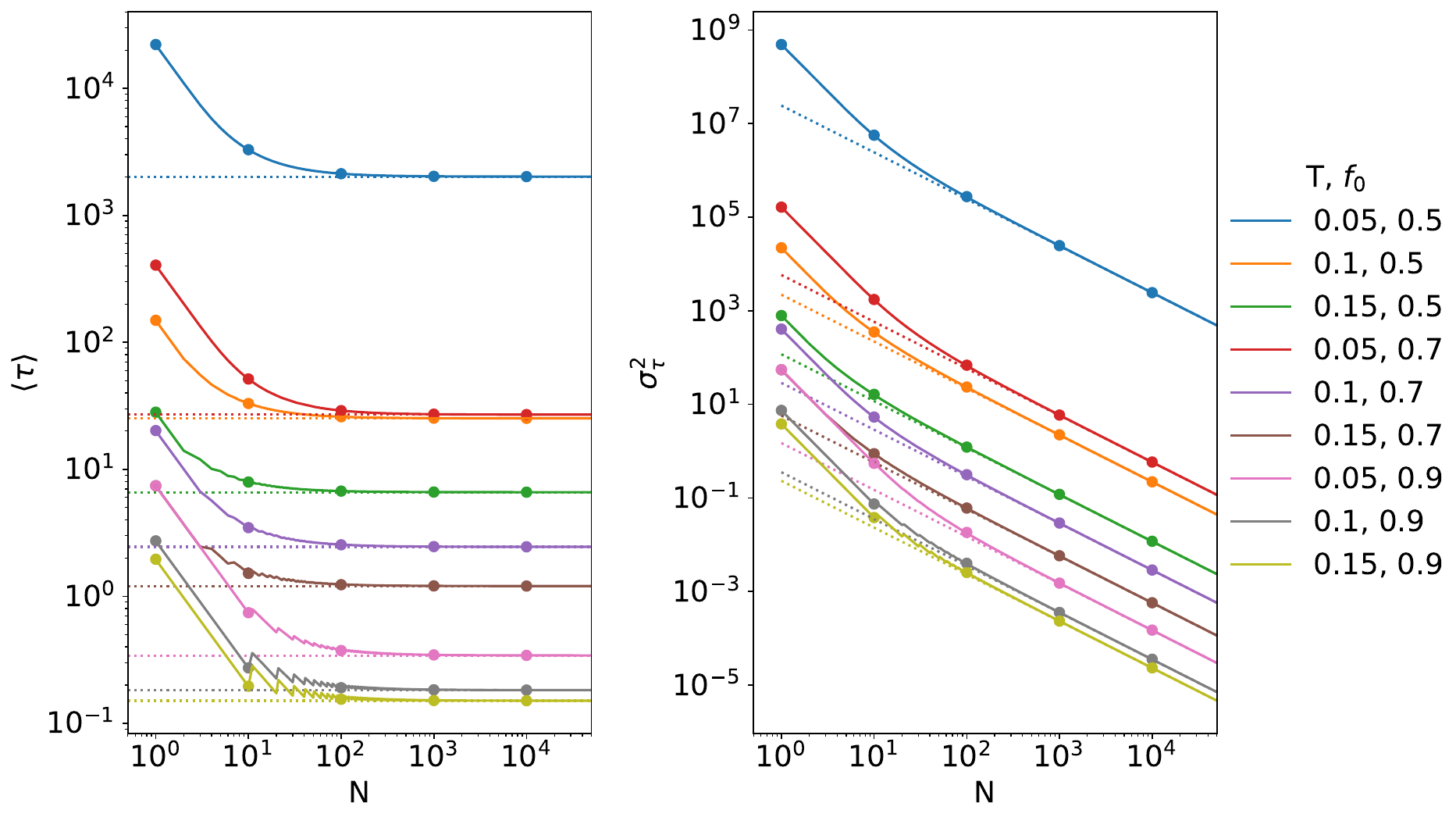}
\caption{Average lifetime (left) and variance (right) for identical thresholds. Dots represent simulation results; dotted lines indicate the asymptotic expressions from Eqs. \ref{identical-mean-limit} and \ref{identical-var-limit}, while solid lines show the exact results from Eqs. \ref{identical-mean-exact} and \ref{identical-var-exact}.}
\label{identical-moments}
\end{figure*}

A more detailed presentation of the following considerations can be found in the first author's PhD thesis \cite{hiemer2023predicting}; here we provide a condensed version. If all thresholds $t$ are equal, the sequence leading to failure after application of a load $f_{0} < t$ divides into two steps: i) Initially, fibers fail individually by thermal barrier crossing events, as in this regime the fiber threshold is always larger than the force acting on the fiber. We denote the number of intact fibers before the \(i\)-th thermal event by
\begin{equation}
    n_{i} = N-i+1
\end{equation} 
and the total number of thermal events until global failure by
\begin{equation}
    N_{\rm ev} = {\rm ceil}\left(N\left(1-\frac{f_{0}}{t}\right)\right),
\end{equation}
where \(\rm ceil\) symbolises up-rounding. The number of intact fibers before the last thermal event is $N_{\min} = N - N_{\rm ev} + 1$. ii) Once the number of thermal events has reached $N_{\rm ev}$, all remaining intact fibers break by overloading in a single avalanche. The total rate for the \(i\)-th event is given by
\begin{equation}
    \nu_{i,{\rm tot}} = n_{i} \exp\left(-\frac{t-f}{T}\right).
    \label{homog-rate}
\end{equation}
The lifetime \(\tau\) is here a sum of exponentially distributed variables, hence it follows a hypoexponential distribution \cite{hypoexp}. Expressions for the probability density and cumulative probability of a generic, hypoexponentially distributed random variable are given in Appendix A. After insertion and rearranging, we obtain
\begin{eqnarray}
    p(\tau) &=& \frac{N!}{N_{\rm ev}!} \exp\left[-T^{-1}\left(t - F\sum_{n=N_{\rm min}}^{N}n^{-1}\right)\right] \notag \\ 
    & &\sum_{n=N_{\rm min}}^{N}\left\{ \exp\left[-n\,\exp\left(-\frac{t-\frac{F}{n}}{T}\right)\tau \right] \right.
    \notag \\ 
    & &\left. \prod_{\stackrel{j=N_{\rm min}}{j \neq n}}^{N_{\rm ev}}\left[j\,\exp\left(\frac{F}{jT}\right) -n\,\exp\left(\frac{F}{nT}\right)\right]^{-1}\right\} \label{identical-pdf} \\
    P(\tau) &=& 1-\frac{N!}{N_{\rm ev}!} \exp\left[\frac{F}{T}\sum_{n=N_{\rm min}}^{N}n^{-1}\right] \notag \\ 
    & & \sum_{n=N_{\rm min}}^{N}\left\{n^{-1} \exp\left[-n\,\exp\left(-\frac{t-\frac{F}{n}}{T}\right)\tau-\frac{F}{nT}\right] \right.\notag \\ & &\left.\prod_{\stackrel{j=N_{\rm min}}{j \neq n}}^{N_{\rm ev}}\left[j\,\exp(\frac{F}{jT}) -n\,\exp(\frac{F}{nT})\right]^{-1}\right\}.
    \label{identical-cdf}
\end{eqnarray} 
The mean $\langle \tau \rangle$ and variance $\sigma^2_{\tau}$ of this distribution can be evaluated directly, using the relations $\langle \tau \rangle = \sum_{i=1}^{N_{\rm ev}} \nu_{i,\rm tot}^{-1}$ and $\sigma^2_{\tau}= \sum_{i=1}^{N_{\rm ev}} \nu_{i,\rm tot}^{-2}$. We obtain
\begin{equation}
    \langle \tau \rangle = \exp\left(\frac{t}{T}\right) \sum_{n=N_{\rm min}}^{N} n^{-1}\exp\left(-\frac{F}{nT}\right)
    \label{identical-mean-exact}
\end{equation}
\begin{equation}
    \sigma^2_{\tau}= \exp\left(\frac{2t}{T}\right) \sum_{n=N_{\rm min}}^{N} n^{-2}\exp\left(-\frac{2F}{nT}\right)
    \label{identical-var-exact}
\end{equation}
In the limit \(N\gg 1\) , the number of events simplifies to \(N_{\rm ev} \approx N(1-\frac{f_{0}}{t})\) and the summations can be converted to integrals:
\begin{eqnarray}
    & \langle \tau \rangle \approx \exp\left(\frac{t}{T}\right) \int_{\frac{F+t}{t}}^{N} n^{-1}\exp\left(-\frac{F}{nT}\right) {\rm d}n \notag \\
    & = \exp\left(\frac{t}{T}\right)\left[
    {\rm Ei}\left(-\frac{Ft}{(F+t)T}\right)
    -{\rm Ei}\left(-\frac{f_{0}}{T}\right)\right], \\
    & \sigma^2_{\tau} \approx \exp\left(\frac{2t}{T}\right) \int_{\frac{F+t}{t}}^{N} n^{-2}\exp\left(-\frac{2F}{nT}\right) {\rm d}n \notag \\
    & = \frac{T}{2Nf_{0}}\,\exp\left(\frac{2t}{T}\right)\left[
    \exp\left(-\frac{2f_{0}}{T}\right)-\exp\left(-\frac{2Ft}{(F+t)T}\right)\right].
\end{eqnarray} 
\({\rm Ei}(x)\) is the exponential integral \cite{table_func}. Further simplifications can be made in the limit \(F\gg t\):
\begin{eqnarray}
    \langle \tau \rangle &\approx& \exp\left(\frac{t}{T}\right)\left[{\rm Ei}\left(-\frac{t}{T}\right)-{\rm Ei}\left(-\frac{f_{0}}{T}\right)\right],
    \label{identical-mean-limit}\\
    \sigma^2_{\tau} &\approx& \frac{T}{2Nf_{0}}\exp\left(\frac{2t}{T}\right)\left[\exp\left(-\frac{2f_{0}}{T}\right)-\exp\left(-\frac{2t}{T}\right)\right]
    \label{identical-var-limit}.\nonumber\\
\end{eqnarray}
As can be seen in Fig. \ref{identical-moments}, the exact expressions and asymptotic limits we derived describe the results of simulations very well. Convergence to the asymptotic limit is often already achieved at roughly one hundred fibers. We note that in the low-temperature limit $T \to 0$ we can use the asymptotic expansion of the exponential integral for large arguments to write approximately 
\begin{eqnarray}
    \langle \tau \rangle_{T \to 0} &\approx& \frac{T}{f_{0}}\exp\left(\frac{t-f_{0}}{T}\right)
    \label{identical-mean-limit-lowT}\\
    \sigma^2_{\tau,T \to 0} &\approx& \frac{T}{2Nf_{0}}\exp\left(\frac{2(t-f_{0})}{T}\right)
    \label{identical-var-limit-lowT}.
\end{eqnarray}

\subsection{Comparison with the GSC model}
The assumption of the model established by Guarino et al.\cite{guarino} and further developed by Roux\cite{roux} is a Gaussian noise \(\eta\) with zero mean and variance \(kT^{*}\) acting on each fiber independently in addition to the static load \(f\) \cite{guarino,roux}. After some algebra Roux arrives at the total rate for the \(i\)-th event
\begin{equation}
    \nu_{i,{\rm tot}}=(-1)(N-i+1)\ln\left[1-P_{\eta}\left(1-\frac{N\frac{f_{0}}{t}}{N-i+1}\right)\right].
    \label{roux-rate}
\end{equation}
where \(P(\eta)\) is the cumulative distribution of the Gaussian noise $\eta$. Again, we are dealing with a total lifetime that is a sum of exponentially distributed variables, thus it follows a hypoexponential distribution. However, the event rates in the GSC model differ substantially from those in the present model, reflecting the different underlying assumptions. In the present model, failure is the result of the cumulative action of many thermal fluctuations, following the classical theory of crossing an energy barrier. In the GSC model, on the other hand, failure occurs if the applied force, plus a {\em single} value of the fluctuating force in the large-force tail of a Gaussian distribution, exceeds the local threshold. In this model, the asymptotic mean lifetime derives, in the low-temperature limit, as
\begin{align}
    \langle \tau \rangle \approx \frac{\sqrt{2\pi kT}} {f_{0}}\exp\left(-\frac{(t-f_{0})^2}{2 k_{\rm B} t}\right).
\end{align}
If one compares the results by Roux for \(\langle\tau\rangle\) and $\sigma^2_{\tau}$  in the limits \(N>>\) and \(kT <<(1-f_{0})^{2}\) with equations \ref{identical-mean-limit} and \ref{identical-var-exact}, one finds that the mean lifetime \(\langle\tau\rangle\) in both cases is independent of \(N\) and the variance also scales in the same manner as \(N^{-1}\). The dependency of the lifetime on temperature and force is, however, different which is not surprising given the different underlying physical assumptions regarding the nature of the thermally activated failure event. 
\section{Fibers with statistically distributed thresholds}
\subsection{Exact  Lifetime Distribution}
\begin{figure}
\includegraphics[width=\columnwidth,height=0.35\textheight,keepaspectratio]{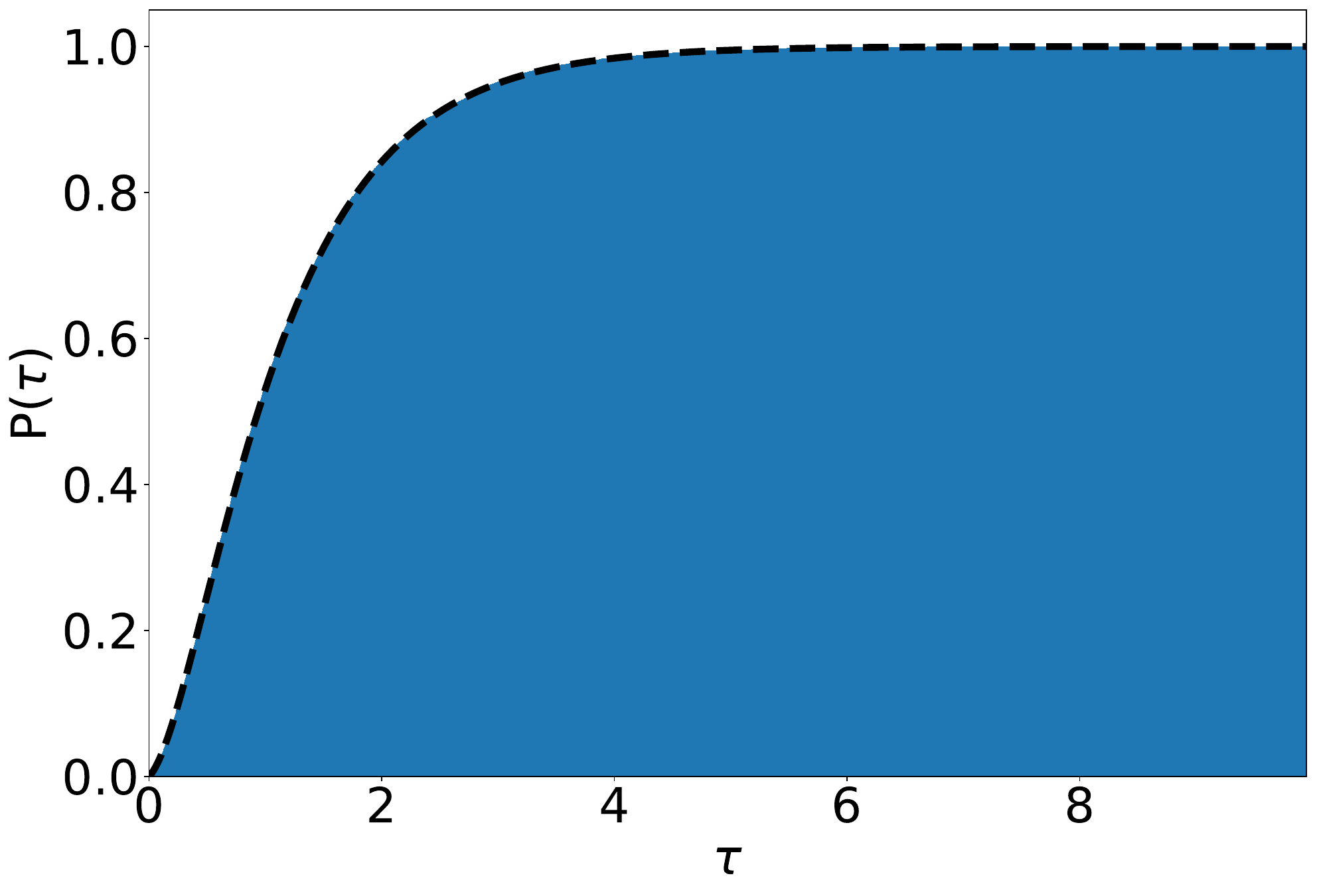}
\caption{Cumulative lifetime distribution for ten fibers with a load of \(0.175\) and temperature in proper units of \(0.15\). Blue is the empirical distribution of 1000 samplings of the same thresholds realization. The black line is the theoretical cumulative distribution function from equation \ref{cdf-lifetime-disordered-realization}.}
\label{plot-lifetime-disordered-realization}
\end{figure}

\begin{figure}
\includegraphics[width=\columnwidth,height=0.35\textheight,keepaspectratio]{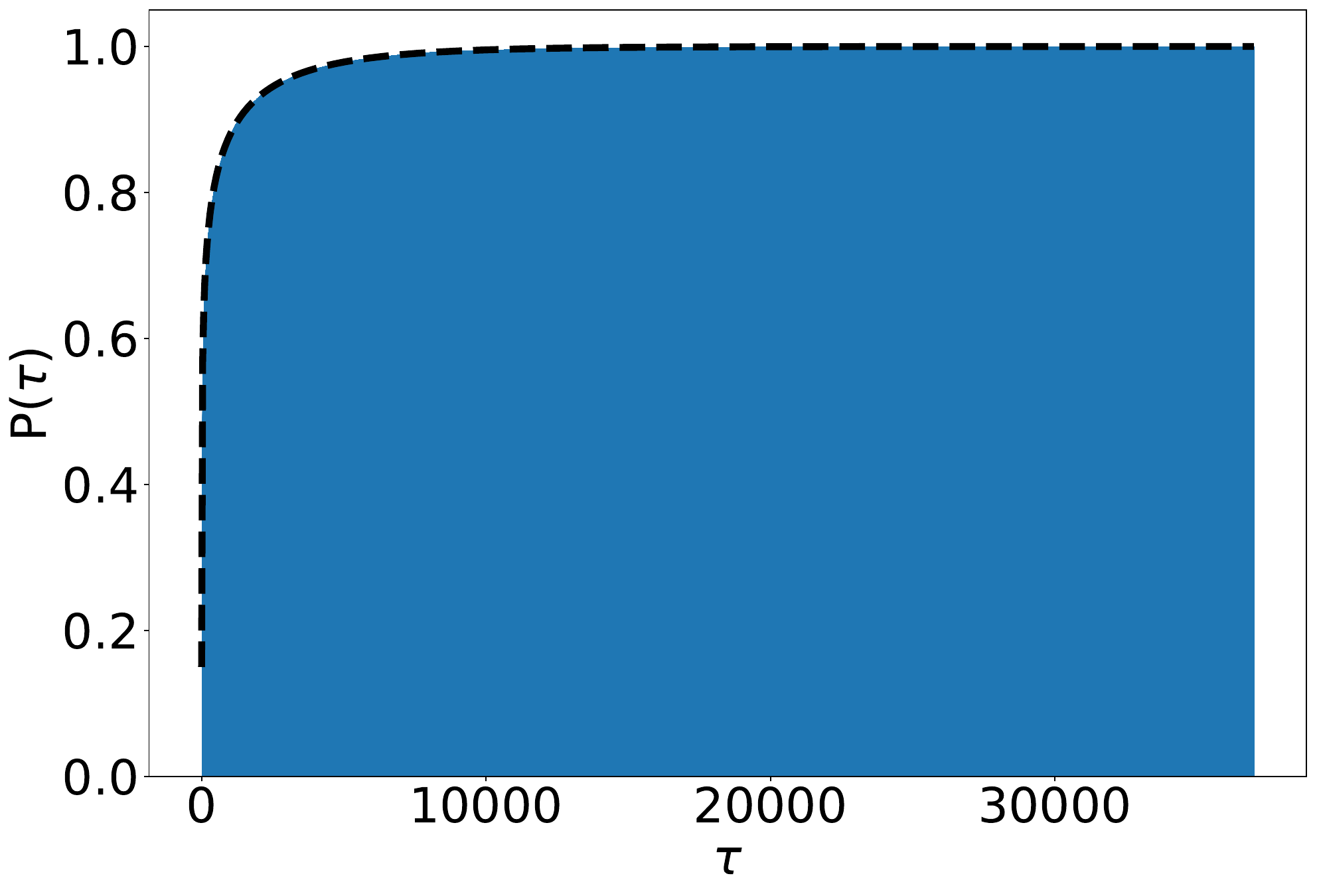}
\caption{Comparison of equation \ref{single-fiber-uniform} with 100000 simulations for temperature 0.1 and load 0.15 for a single fiber and a uniform distribution. The dashed black line is the theoretical prediction whereas the blue color is the empirical histogram.}
\label{single-fiber-plot}
\end{figure}

For random fiber thresholds following the distribution \(p(t)\), the analysis becomes considerably more complex, as the total failure rate \(\nu_{\text{tot}}\) depends on the particular fibers that fail due to individual thermal events. Each such event changes the set of intact fibers and thus modifies the overall failure dynamics. We therefore break the derivation into a sequence of easier subproblems:
\begin{enumerate}
\renewcommand{\labelenumi}{\roman{enumi})}
    \item given a fiber bundle of size \(N\) with known thresholds \(\boldsymbol{t}\), an initial force per fiber \(f_{0}\), and a known failure sequence \(S\) of thermally activated failures, what is \(p\left(\tau | S, \boldsymbol{t}\right)\) (lifetime distribution conditioned on \(S\) and \(\boldsymbol{t}\))?
    \item assuming \(p(\tau|S, \boldsymbol{t})\) is known, what is the probability of the specific failure sequence \(S\) given the thresholds \(\boldsymbol{t}\), i.e., what is \(P(S | \boldsymbol{t})\)?
    \item Given \(p(\tau|S,\boldsymbol{t})\) and \(P(S | \boldsymbol{t})\), what is the lifetime distribution conditioned on the threshold configuration \(\boldsymbol{t}\), \(p(\tau | \boldsymbol{t})\)?
    \item Given \(p(\tau|\boldsymbol{t})\) and the threshold distribution \(p(t)\), what is the overall lifetime distribution \(p(\tau)\)?
\end{enumerate} 
Starting with sub-problem i), we realize that in the case of a known failure sequence \(S\) and \(\boldsymbol{t}\), the lifetime \(\tau\) is a sum of exponentially distributed variables as in the homogeneous fiber bundle model. Therefore,\(p\left(\tau | S, \boldsymbol{t}\right)\) follows a hypoexponential distribution. Passing on to sub-problem ii), the probability of \(S\) given \(\boldsymbol{t}\) can be taken from Eq. \ref{prob_chosen}
\begin{equation}
    P(S|\boldsymbol{t})=\prod_{i=1}^{N_{\rm ev}^{S}}\frac{ \nu_{i,{\rm q}}}{\nu_{i,{\rm tot}}}
\end{equation}
where \(\nu_{i,{\rm q}}\) is the failure rate of the fiber \(q\) that fails through thermal activation, \(\nu_{i,{\rm tot}}\) the total rate at the \(i\)-th event and \(N_{\rm ev}^{S}\) the number of events of the respective sequence of failures \(S\). The conditional lifetime distribution \(p(\tau|\boldsymbol{t})\) is then sum over all lifetime distributions of individual failure sequences \(p(\tau|S,\boldsymbol{t})\) weighted by the failure sequence probability \(P(S|\boldsymbol{t})\)
\begin{equation}
    p(\tau|\boldsymbol{t})=\sum_{S \in \mathcal{S}_{\boldsymbol{t}}} P(S|\boldsymbol{t})p(\tau|S,\boldsymbol{t})
\end{equation}
thus solving sub-problem iii). \(\mathcal{S}_{\boldsymbol{t}}\) denotes the set of all physically viable failure sequences given the set of thresholds \(\boldsymbol{t}\). A nonphysical failure sequence would be e. g. if two fibers fail thermally at precisely the same time, a fiber fails that has already failed earlier, a fiber fails critically that is stronger than the current force \(f\), etc. pp. In other words a non-physical sequence is in contradiction with the assumptions of the model. The full expression after rearranging of terms then reads as
\begin{align}
    p(\tau|\boldsymbol{t})= &\sum_{S \in \mathcal{S}_{\boldsymbol{t}}}\left\{ \left(\prod_{i=1}^{N_{\rm ev}^{S}}\nu_{i,{\rm q}}\right) \, \sum_{i=1}^{N_{\rm ev}^{S}} \left[ \exp(-\nu_{i,{\rm tot}}\tau) \right.\right.
    \nonumber \\ & \left.\left. \prod_{j=1, j \neq i}^{N_{\rm ev}^S}(\nu_{j,{\rm tot}}-\nu_{i,{\rm tot}})^{-1}\right]\right\} 
    \label{lifetime-disordered-realization}\\
    P(\tau|\boldsymbol{t})= &\sum_{S \in \mathcal{S}_{\boldsymbol{t}}}\left\{ \left(\prod_{i=1}^{N_{\rm ev}^{S}}\nu_{i,{\rm q}}\right) \, \sum_{i=1}^{N_{\rm ev}^{S}}[1-\frac{\exp(-\nu_{i,{\rm tot}}\tau)}{\nu_{i,{\rm tot}}} \right. \nonumber \\ &\left.\prod_{j=1, j \neq i}^{N_{\rm ev}^S}(\nu_{j,{\rm tot}}-\nu_{i,{\rm tot}})^{-1}]\right\}.
    \label{cdf-lifetime-disordered-realization}
\end{align}
\(N_{\rm ev}^{S}\) is the number of events within a specific failure sequence \(S\). Distributions of this kind are called phase type distributions and are common in queuing theory \cite{phase}. 
As an intermediate check for correctness of the calculations so far, we compare with simulations where the stochastic creep dynamics are simulated repeatedly on the same realization of thresholds. The simulation data displayed in Fig. \ref{plot-lifetime-disordered-realization} and Eq. \ref{cdf-lifetime-disordered-realization} are in perfect agreement.
We now proceed to eliminate the remaining conditioning on \(\boldsymbol{t}\) from \(p(\tau|\boldsymbol{t})\), ergo subproblem iv): Eqs. \ref{lifetime-disordered-realization} and \ref{cdf-lifetime-disordered-realization} applies to all threshold configurations \(\boldsymbol{t}\) that share the same \(\mathcal{S}_{\boldsymbol{t}}\). We recall that the applied forces in the fiber bundle model in creep are limited to discrete values ("quantized") \(f_{k}\)
\begin{equation}
    f_{k} = \frac{F}{N-k+1}
    \label{force_ind}
\end{equation}
(note that with this convention \(f_{0}=f_{1}\)). So which \(\boldsymbol{t}\) share the same set \(\mathcal{S}_{\boldsymbol{t}}\)? 
For answering this question, it is advantageous to consider the low temperature limit: a big initial avalanche without thermal activation is triggered initially when the force is applied. This avalanche eliminates the \(a_0\) weakest fibers within the bundle. Subsequently, all fibers fail either thermally in order of the size of the threshold or critically due to load-distribution from another fiber failure, until the entire fiber bundle collapses in the final big avalanche. \\
Therefore as a first step, we order the elements of \(\boldsymbol{t}\) from weakest to strongest \footnote{(so \(t_{(1)}\) is the weakest)} and exclude the \(a_{0}\) weakest fibers from consideration as they do not alter \(\mathcal{S}_{\boldsymbol{t}}\) and do not change the temporal dynamics of the fiber bundle. As a result of the ordering we switch from the threshold distribution \(p(t)\) to its order statistics where we denote the \(i\)-th order statistics by \(p(t_{(i)})\). As conditions we note that the first \(a_{0}\) thresholds in the order statistics must be smaller than \(f_{a_{0}}\) to fail collectively and the largest \(N-a_{0}\) thresholds in the order statistics have to be larger than \(f_{a_{0}+1}\) as otherwise they become part of the initial avalanche \(a_{0}\). \\
As second step, the number of critically failing fibers \(a_\infty\) in the final avalanche causing global failure can be identified by the considering the largest order statistics. Forming an ordered subset from the \((N-a_\infty)\)-th order up to the \(N\)-th order statistic/maximum, every member \(k\) of this group has to be smaller than \(f_{N-a_{\infty}+k}\) in order to fail in the last avalanche, but larger than \(f_{N-a_{\infty}-1}\) in order to fail \textit{just} in the last avalanche and not before. This places a condition on the \((N-a_\infty-1)\)-th order statistic to be in the interval \( f_{N-a_\infty} < t_{(-a_\infty-1)} \leq f_{N-a_\infty+1}\). This means that the \((N-a_\infty-1)\)-th strongest fiber can just bear the load if all stronger fibers are still intact, but not after one of them has failed. 
As third and final step, we state that two realizations \(\boldsymbol{t}\) share the same \(\mathcal{S}_{\boldsymbol{t}}\), if every order statistic \(i\) with \(a_{0}<i<N-a_{\infty}\) is within the same interval 
\begin{equation}
    f_{c_{i}+1} < t_{(i)} \leq f_{c_{i}}.
\end{equation}
\(\boldsymbol{c}\) is a monotonously increasing sequence of \(N-a_{0}-a_{\infty}-1\) integers with values ranging from \(a_{0}+1\) to \(N-a_{\infty}\). It can be interpreted as an integer label for each order statistic, specifying the total number of preceding fiber failures required for that fiber to reach its failure threshold.
The lower bound \(a_{0}+1\) is necessary to avoid any of the order statistics to fall within the first avalanche \(a_{0}\) and the upper bound follows from the condition on the \((N-a_\infty-1)\)-th order statistics.
We can now transform the conditioning on \(\boldsymbol{t}\) to a conditioning on the set of integers \(a_{0},a_{\infty},\boldsymbol{c}\) by weighting the thresholds by the joint probability density of the \(N-a_{0}\) largest order statistics \(p\left(t_{(a_{0}+1)},t_{(a_{0}+2)},...,t_{(N)}\right)\) and integrating within the previously discussed bounds:
\begin{align}
    p(\tau|a_{0},\boldsymbol{c},a_{\infty})&= C_{N}\int_{f_{c_{1}}}^{f_{c_{1}+1}}\int_{f_{c_{2}}\text{ or }t_{(a_{0}+1)}}^{f_{c_{2}+1}}... \notag \\ 
    & \int_{f_{c_{N-a_{0}-a_{\infty}-1}}\text{ or }t_{(N-a_{\infty}-3)}}^{f_{c_{N-a_{0}-a_{\infty}-1}+1}} \notag \\ 
    & \int_{f_{N-a_{\infty}}\text{ or }t_{(N-a_{\infty}-2)}}^{f_{N-a_{\infty}+1}}\int_{t_{(N-a_{\infty}-1)}}^{f_{N-a_{\infty}+1}}... \notag \\ 
    & \int_{t_{(N-1)}}^{f_{N}} p\left(t_{(a_{0}+1)},t_{(a_{0}+2)},...,t_{(N)}\right) \notag \\ 
    & \sum_{S \in \mathcal{S}_{\boldsymbol{t}}}\{ (\prod_{i=1}^{N_{\rm ev}^{S}}\nu_{i,{\rm q}}) \, \notag \\ &\sum_{i=1}^{N_{\rm ev}^{S}}[ \exp(-\nu_{i,{\rm tot}}\tau)\prod_{j=1, j \neq i}^{N_{\rm ev}^S}(\nu_{j,{\rm tot}}-\nu_{i,{\rm tot}})^{-1}]\} \notag \\ 
    & dt_{(a_{0}+1)},dt_{(a_{0}+2)}...dt_{(N)}.
    \label{lifetime-disordered-categorized}
\end{align}
\(C_{N}\) is a normalization constant, that ensures \(\int_{0}^{\infty}p(\tau|a_{0},\boldsymbol{c},a_{\infty})d\tau=1\). In the lower integration bounds appearing in the previous equation, expressions such as "\(f_{c_{2}}\text{ or }t_{(a_{0}+1)}\)" indicate that the integral bounds need to be adapted if \(c_{i}=c_{i+1}\) to ensure the correct ordering  \(t_{(a_{0}+i)}\leq t_{(a_{0}+i+1)}\). Now by summing over all possible combinations \(a_{0}\),\(a_{\infty}\) and \(\boldsymbol{c}\) one arrives at the final expression for the lifetime distribution:
\begin{equation}
    p(\tau) = \sum_{a_{0}=0}^{N}\sum_{a_{\infty}=0}^{N-a_{0}-1} \sum_{\boldsymbol{c}}P(a_{0})P(\boldsymbol{c},a_{\infty})p(\tau|a_{0},\boldsymbol{c},a_{\infty})
    \label{lifetime-disordered}
\end{equation}
As a simple test to check the validity of the above expression we analyze it for a uniform distribution and a single fiber (equation \ref{single-fiber-uniform}) and compare it to simulations in Fig. \ref{single-fiber-plot} with perfect agreement. 
The probability of an individual combination  \(a_{0},\boldsymbol{c},a_{\infty}\) can be adapted from Daniel's equation for the probability of failure of the entire bundle \cite{daniels}:
\begin{align}
    & P(a_{0},\boldsymbol{c},a_{\infty})=
    \int_{0}^{f_{1}}\int_{t_{(1)}}^{f_{2}}...\int_{t_{(a_{0})}}^{f_{a_{0}}} \int_{f_{c_{1}}}^{f_{c_{1}+1}} \notag \\ 
    & \int_{f_{c_{2}}}^{f_{c_{2}+1}}... \int_{f_{c_{N-a_{0}-a_{\infty}-1}}\text{ or }t_{(N-a_{\infty}-3)}}^{f_{c_{N-a_{0}-a_{\infty}-1}+1}} \notag \\ 
    & \int_{f_{N-a_{\infty}}\text{ or }t_{(N-a_{\infty}-2)}}^{f_{N-a_{\infty}+1}}\int_{t_{(N-a_{\infty}-1)}}^{f_{N-a_{\infty}+1}}... \notag \\ 
    & \int_{t_{(N-1)}}^{F} p\left(t_{(1)},t_{(2)},...,t_{(N)}\right) dt_{(1)},dt_{(2)}...dt_{(N)}
    \label{probabilities} 
\end{align}
\(p\left(t_{(1)},t_{(2)},...,t_{(N)}\right)\) is the joint distribution of all order statistics. This can be readily solved for a few fibers, but results in a sum which has \(2^{N}\) terms where each term in the sum is a product of up to \(N\) different factors. To find the number of combinations over which one has to sum, only combinations leading to unique \(\mathcal{S}_{\boldsymbol{t}}\) have to be considered. The number of combinations of \(a_{0}\) and \(a_{\infty}\) can be found by all combinations that fulfill the inequality \(N \geq a_{0}+a_{\infty}+1 \). The \(+1\) is needed as \(a_{\infty}\) is the number of \textit{critically} failed fibers in the last avalanche. From that we can build a summation that can be reduced via Gauss's summation formula   
\begin{equation}
    1+\sum_{a_{0}=0}^{N-1}(N-a_{0})=1+\frac{N}{2}\left(N+1\right). \notag
\end{equation}
For the full number of combinations, we recall that the values of \(\boldsymbol{c}\) are within the range \([a_{0}+1,N-a_{\infty}]\), \(\boldsymbol{c}\) has \(N-a_{0}-a_{\infty}-1\) entries and is a monotonously increasing sequence. Thus this is equivalent to the number of outcomes of drawing with repetition \((N-a_{0}-a_{\infty}-1)\) times from an urn with \((N-a_{0}-a_{\infty})\) different objects with no regard for order. The result of this is the multiset \(\left(\binom{n}{k}\right)=\binom{n+k-1}{k}\) with \(n=N-a_{0}-a_{\infty}\) elements and size \(k=N-a_{0}-a_{\infty}-1\):
\begin{equation}
    N_{\rm comb}=1+\sum_{a_{0}=0}^{N-1}\sum_{a_{\infty}=0}^{N-a_{0}-1}\frac{ \left[2\left(N-a_{0}-a_{\infty}-1\right)\right]!}{\left(N-a_{0}-a_{\infty}-1\right)!^{2}}
    \label{eq:combinations}
\end{equation} 
As can be easily seen, this term explodes and thus makes the summation in Eq. \ref{lifetime-disordered} hard to evaluate even for few fibers and impossible for large fibers. Future research should thus be directed towards finding the asymptotic form of Eq. \ref{lifetime-disordered}.

\subsection{Simulation Results}
Fiber bundle simulations have been conducted with parameters \(1\leq N\leq10^{5}\), \(\frac{f_{0}}{f_{c}}=[0.5,0.6,0.7]\) and \(T=[0.05,0.1,0.15]\). \(f_{c}\) is the theoretical critical force which leads to immediate catastrophic breakdown. For a uniform threshold distribution of unit range, \(f_{c}\) is \(0.25\)  whereas for an exponential threshold distribution with unit average it is \(e^{-1}\) \cite{alava}. 
\subsubsection{Average and Variance of the Lifetimes}
\begin{figure*}
\includegraphics[width=\textwidth,height=0.32\textheight,keepaspectratio]{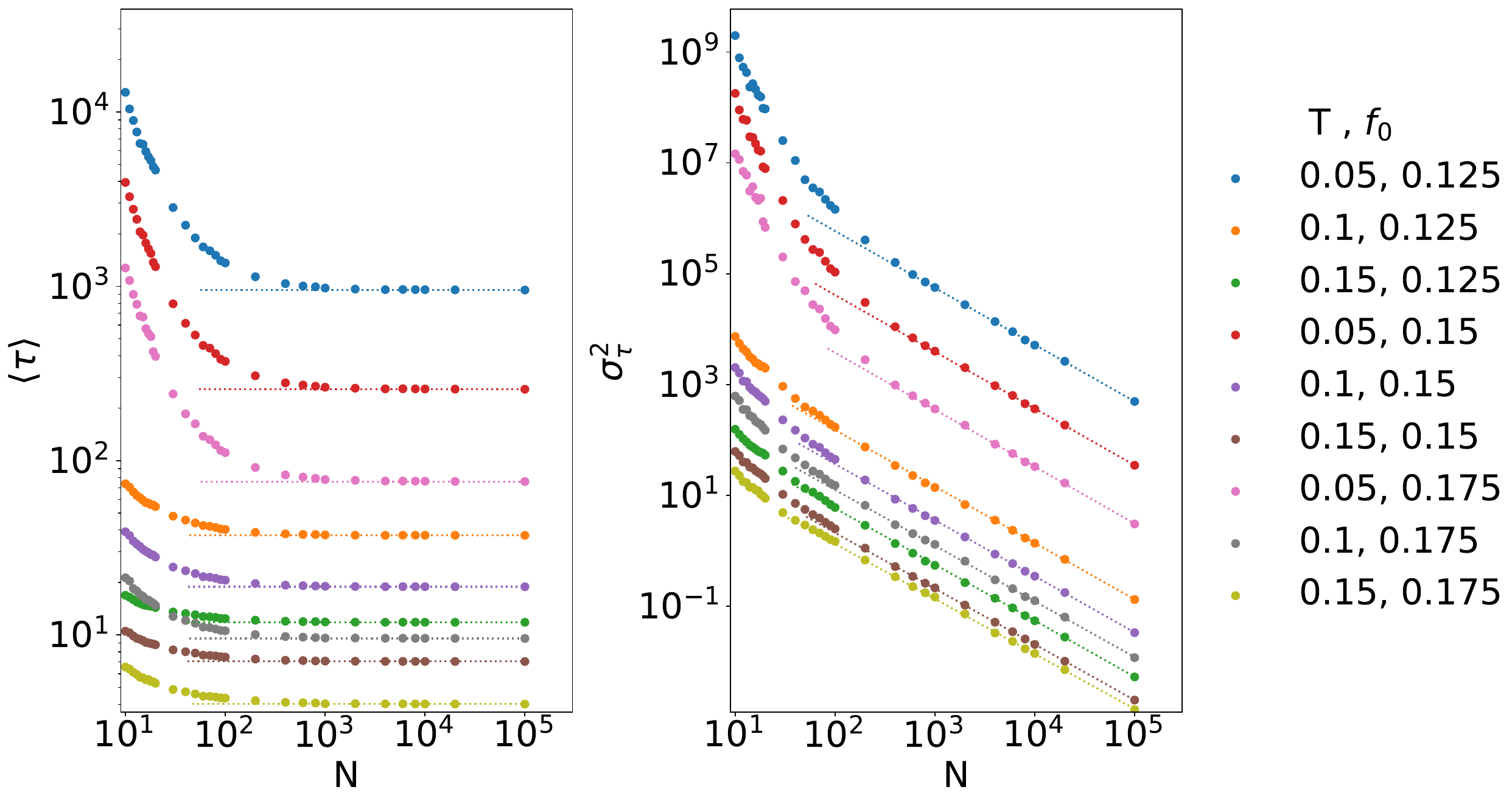}
\caption{Average lifetime and variance for thresholds drawn from a uniform distribution spanning from zero to one as a function of bundle size $N$. The dotted line is the asymptotic limit which for the average is the taken as the value of the largest bundle size which is extended by a horizontal line.}
\label{moment-disordered-uniform}
\end{figure*}

\begin{figure*}
\includegraphics[width=\textwidth,height=0.32\textheight,keepaspectratio]{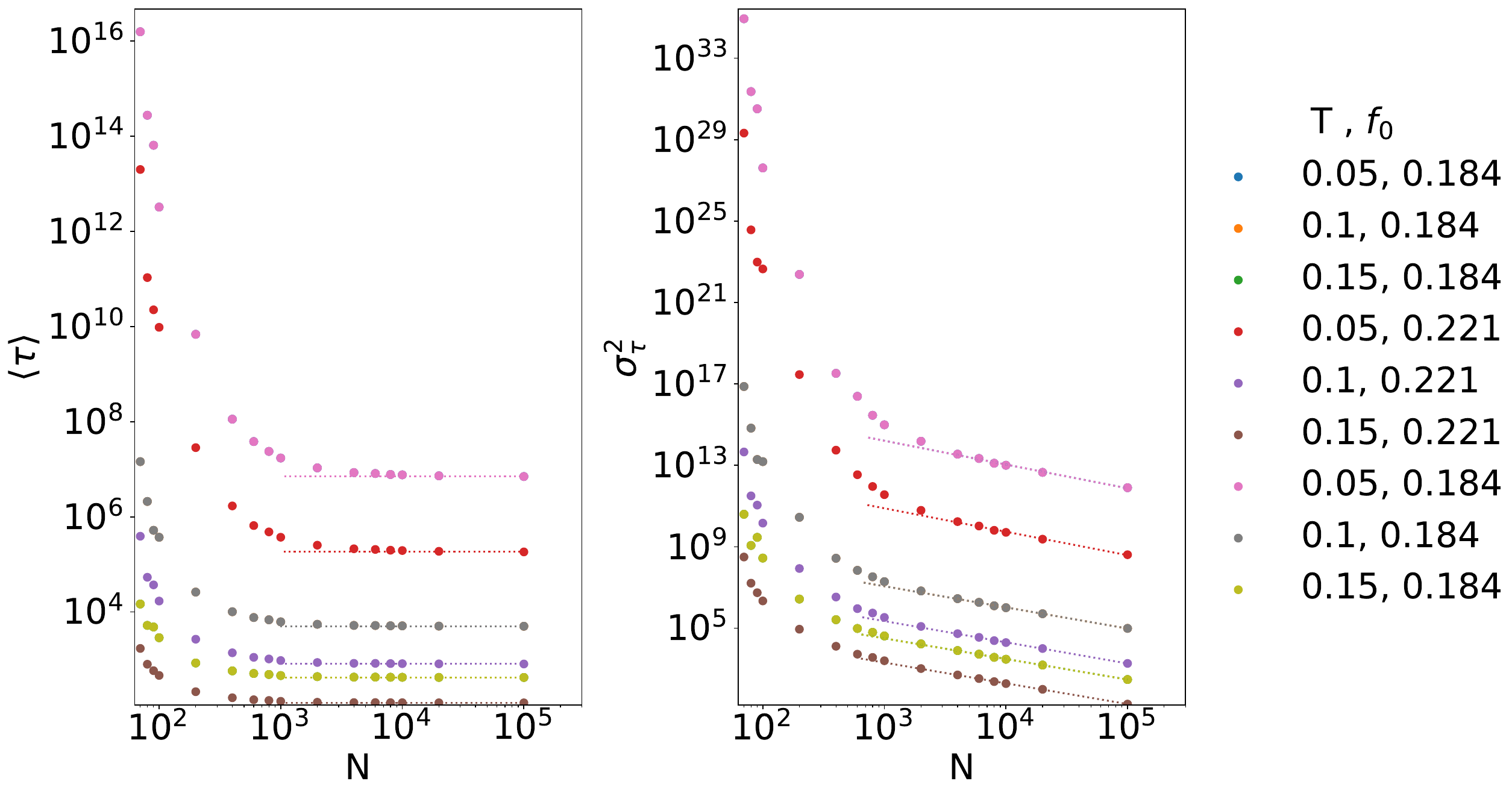}
\caption{Average lifetime and variance for thresholds drawn from an exponential distribution $p\left(t\right) = \exp\left(t\right)$, as a function of bundle size $N$. The last value is taken as asymptotic limit and extended with the dotted line. The dotted line is the asymptotic limit which for the average is the taken as the value of the largest bundle size which is extended by a horizontal line.}
\label{moment-disordered-exp}
\end{figure*}

The simulation results for average lifetime and lifetime variance obtained for uniformly and exponentially distributed thresholds can be seen in Figures \ref{moment-disordered-uniform} and \ref{moment-disordered-exp}.  
The average lifetime approaches a constant asymptotic value for large fiber numbers, regardless of the threshold distribution, and the variance shows an asymptotic power law decay with exponent \(n_{var}\approx1\). The specific values for the asymptotic mean lifetime \(\langle\tau_{\infty}\rangle\) and variance scaling exponent can be seen in Tab. \ref{table-lifetime-fitparams}. 
The asymptotic behaviour for statistically distributed fiber thresholds in the thermodynamic limit is thus qualitatively the same as for a single threshold and appears not to be affected by the details of the threshold distribution at least within the limited range of cases considered here. 


\begin{table}[b]
\caption{Overview over the fitting parameters of the asymptotic limits for the average and the variance of lifetimes. \(\langle \tau_{\infty}\rangle\) is the average lifetime value at 100000 fibers as proxy for the thermodynamic limit and \(n_{var}\) is the power law exponent of the variance.}
\label{table-lifetime-fitparams}
\begin{ruledtabular}
\begin{tabular}{ccccc}
\(f_{0}\) & \(T\) & \(\langle \tau_{\infty}\rangle\) & \(n_{var}\) & threshold distribution \\
\hline
0.125 & 0.05 & 952.965 & 1.016 & uniform\\
0.125 & 0.1 & 37.307 & 1.013 & uniform\\
0.125 & 0.15 & 11.828 & 1.009 & uniform\\
0.15 & 0.05 & 256.949 & 1.015 & uniform\\
0.15 & 0.1 & 18.922 & 1.007 & uniform\\
0.15 & 0.15 & 7.046 & 0.999 & uniform\\
0.175 & 0.05 & 75.946 & 1.011 & uniform\\
0.175 & 0.1 & 9.561 & 0.987 & uniform\\
0.175 & 0.15 & 4.015 & 0.986 & uniform\\
0.184 & 0.05 & 7035759.5 & 1.169 & exponential\\
0.184 & 0.1 & 5006.723 & 1.039 & exponential\\
0.184 & 0.15 & 419.75 & 1.012 & exponential\\
0.221 & 0.05 & 183014.938 & 1.151 & exponential\\
0.221 & 0.1 & 806.157 & 1.041 & exponential\\
0.221 & 0.15 & 122.938 & 1.019 & exponential\\
0.258 & 0.05 & 8361.321 & 1.126 & exponential\\
0.258 & 0.1 & 170.477 & 1.036 & exponential\\
0.258 & 0.15 & 42.232 & 1.017 & exponential\\
\end{tabular}
\end{ruledtabular} 
\end{table}
\subsubsection{Avalanche Size Distribution}
\begin{figure}
\includegraphics[width=\columnwidth]{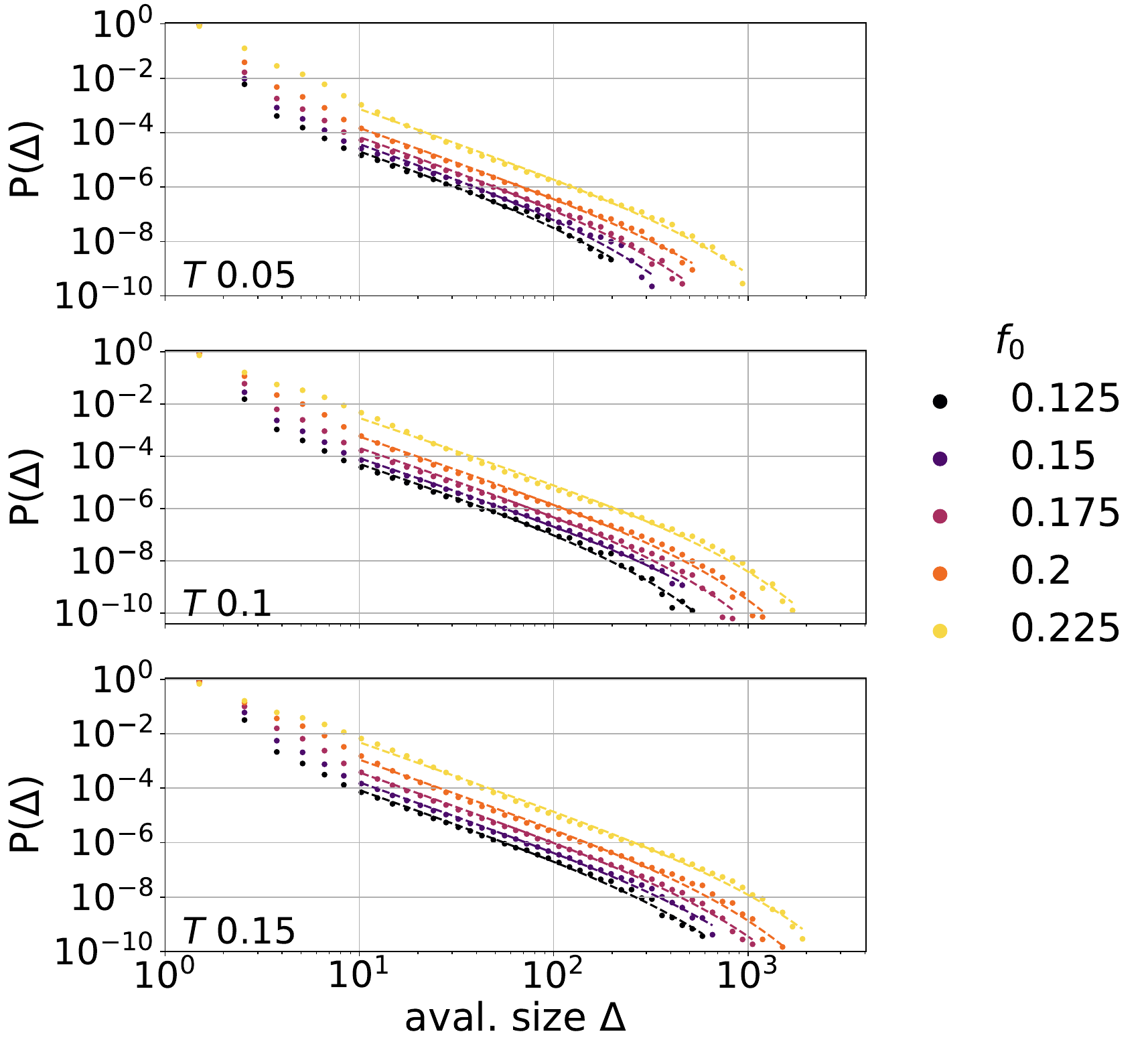}
\caption{Logarithmically binned avalanche distributions (bin size 1.125) after exclusion of the first and last avalanche (\(a_{0}\) and \(a_{\infty}\)) from 1000 independent bundles of \(10^{5}\) fibers with thresholds drawn 
from a uniform distribution spanning from zero to one.} 
\label{avalanche-uniform-individ-plot-scatter}
\end{figure}

\begin{figure}
\includegraphics[width=\columnwidth]{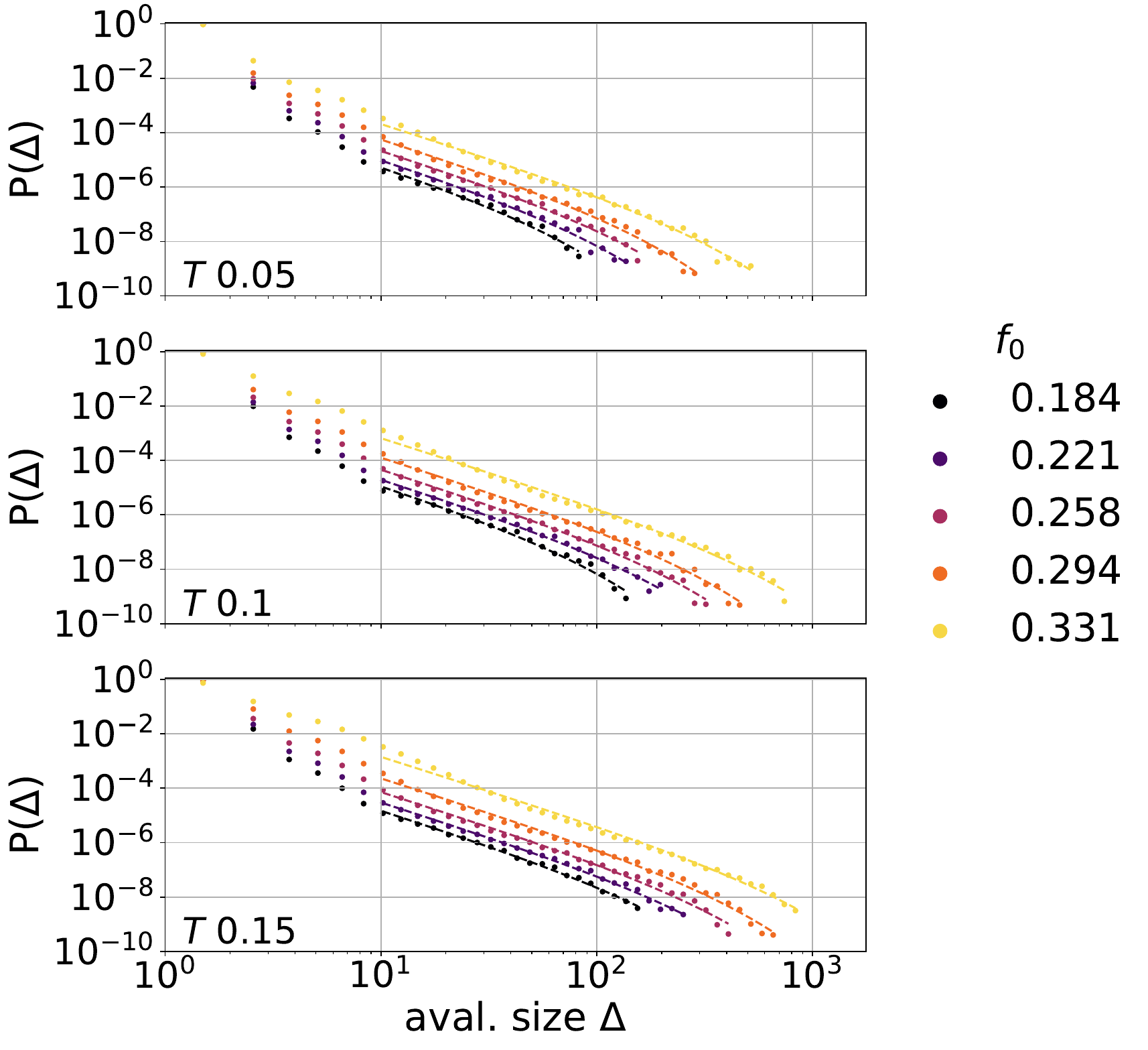}
\caption{Logarithmically binned avalanche distributions (bin size 1.125) after exclusion of the first and last avalanche (\(a_{0}\) and \(a_{\infty}\)) from 1000 independent bundles of \(10^{5}\) fibers with thresholds drawn 
from an exponential distribution of unit mean. }
\label{avalanche-exponential-individ-plot-scatter}
\end{figure}

To investigate the avalanche distribution in the limit \(N\gg 1\), avalanche size distributions of bundles of 100000 fibers with exponentially and uniform distributed thresholds have been investigated. Simulation results with fits of form 
\(\Delta_{0}\exp\left(-c \Delta\right)\Delta^{-n}\)
are shown in Fig.  \ref{avalanche-uniform-individ-plot-scatter} and \ref{avalanche-exponential-individ-plot-scatter}. We find asymptotic power law behaviour with exponent \(\approx2.5\) in the regime of large avalanches. The latter finding agrees well with Hemmer's exponent found for quasistatic loading  of fiber bundles \cite{hemmer}. This agreement can be qualitatively explained by making two observations: i) The most likely fracture path is identical with Hemmer's path and thus yields the same avalanches ii) Hemmer's arguments for the quasistatic case loosely paraphrased are: i) for an avalanche of size \(\Delta\), \(\Delta-1\) fibers must fail via immediate load redistribution. ii) fibers which have already failed are weaker than the weakest fiber failing in the current avalanche. Observation i) is still true in our model while ii) is not, as in principle  in our model any fiber can fail by a thermal barrier-crossing event before the load exceeds the threshold. However, the fraction of fibers significantly violating ii) should be negligible in the low-temperature limit as the probability of a fiber failing via a thermal event decreases exponentially with \(t/T\). 

\section{Conclusion}
In this work, we studied thermally activated breakdown of a fiber bundle model 
based on transition state theory \cite{eyring}. For identical fiber thresholds, the lifetime distribution, average, variance and their asymptotic limits for many fibers are derived 
The asymptotic limits with respect to the number of fibers \(N\) for the average lifetime is independent of \(N\) and the variance scales  by \(N^{-1}\)  which agree with analytical approximations in the low temperature limit made by Roux for a model of thermal breakdown with stationary Gaussian noise \cite{roux,guarino}. For the case of stochastically distributed fiber strengths, the lifetime distribution is derived, but no closed form solution even for simple threshold distributions could be obtained. \\
Simulation results for disordered system of uniform and exponentially distributed fiber thresholds indicate identical asymptotic behaviour for the average lifetime and variance of the lifetimes. The avalanche distributions in these simulations show an asymptotic power law behaviour with exponent \(\approx2.5\) for the limit \(N>>\) coinciding with Hemmer's asymptotic result for quasistatic loading \cite{hemmer}. 
Interesting directions for future research include generalizations to more complex constitutive behaviors, such as viscoelastic fibers~\cite{hidalgo2002creep} or continuous damage models~\cite{kun2000damage}, as well as alternative load redistribution schemes like local load sharing~\cite{phoenix1983comparison}. Another promising avenue is to explore whether analogies can be drawn with more realistic continuum models of composite materials, such as optimal \(n\)-rank laminates\cite{lurie1984exact,norris1985differential,francfort1986homogenization,milton1986modelling}.
\section*{Data Availability statement}
The python scripts used to create the simulations are attached as electronic supplement and can be found \url{https://github.com/ComplexityBiosystems/fiber-bundle}. The research data can be found on Zenodo with DOI 10.5281/zenodo.16413747.
\begin{acknowledgments}
This research was funded by the Deutsche Forschungsgemeinschaft (DFG, German Research Foundation) - 377472739/GRK 2423/2-2023 and via the Grant no. 1 ZA  171/14-1. SH also acknowledges the support by the Humboldt foundation through the Feodor-Lynen fellowship during the later stages of the work.
\end{acknowledgments}

\appendix

\section{Hypoexponential distribution}

We consider a generic variable $x = \sum_{i=1}^k x_i$ which is the sum of $k$ exponentially distributed random variables with cumulative distribution functions $P(x_i) = 1 - \exp(-\lambda_i x_i)$. The probability density $p(x)$ is then given by
\begin{equation}
    p(x)= \sum_{i=1}^{k}[\lambda_i \exp(-\lambda_i x)\prod_{j=1, j \neq i}^{k}\frac{\lambda_j}{\lambda_j-\lambda_i}] 
\end{equation}
and the cumulative distribution function is 
\begin{equation}
    P(x)= 1-\sum_{i=1}^{k}[ \exp(-\lambda_i x)\prod_{j=1, j \neq i}^{k}\frac{\lambda_j}{\lambda_j-\lambda_i}] 
\end{equation}
The mean value of $x$ is given by
\begin{equation}
    \langle x \rangle = \sum_{i=1}^{k} \lambda_{i}^{-1}
\end{equation}
amd the variance is
\begin{equation}
    Var(x) = \sum_{i=1}^{k} \lambda_{i}^{-2}.
\end{equation} 

\section{Single Fiber}
\subsection{Single fiber: General probability distribution}
For a single fiber with the threshold probability density \(p\left(t\right)\) to derive its lifetime, the cases of immediate failure \(f_{0}\geq t\) and thermal failure \(f_{0}<t\) have to be distinguished 
\begin{equation}
    p(\tau|f_{0}\geq t)=\delta(\tau)
\end{equation}
\begin{align}
    p(\tau|f_{0} < t)&=C_{N}\int_{f_{0}}^{\infty}p\left(t\right)\nu \exp(-\nu\tau)dt \\
    C_{N} &= \int_{0}^{\infty} p(\tau|f_{0} < t) d\tau \notag 
\end{align}
The lifetime distribution then reads as
\begin{align}
    p(\tau)&=P(0\leq t \leq f_{0}) p(\tau|f_{0}\geq t) + P(f_{0}\leq t \leq \infty)  p(\tau|f_{0} < t).
\end{align}
\subsection{Single Fiber: Uniform distribution}
We assume the uniform distribution to span from zero to \(\lambda\). Plugging the probability density function of the uniform distribution into the equations of the previous section yields the normalization constant
\begin{equation}
    C_{N}=(\lambda-f_{0})^{-1},
\end{equation}
the conditional lifetime probability density function for the case of thermal failure 
\begin{equation}
    p(\tau|f_{0}<t)=\frac{T}{\lambda-f_{0}}\tau^{-1}\{\exp[-\exp(-\frac{\lambda-f_{0}}{T})\tau]-\exp[-\tau]\}
\end{equation}
and the total lifetime probability density function
\begin{equation}
    p(\tau)=\frac{f_{0}}{\lambda}\delta(\tau) + \frac{T}{\lambda}\tau^{-1}\{\exp[-\exp(-\frac{\lambda-f_{0}}{T})\tau]-\exp[-\tau]\}
\end{equation}
from which we extract the cumulative lifetime distribution via integration over \(\tau\)
\begin{equation}
    P(\tau)=1+\frac{T}{\lambda}\{Ei[-\exp(-\frac{\lambda-f_{0}}{T})\tau]-Ei(-\tau)\}.
    \label{single-fiber-uniform}
\end{equation}  
\subsection{Single Fiber: Exponential distribution}
We assume the exponential distribution with rate \(\lambda\). Plugging the probability density function of the uniform distribution into the equations of the previous section yields the normalization constant
\begin{equation}
    C_{N}=(-1)\lambda e^{f_{0}\lambda},
\end{equation}
the conditional lifetime probability density function for the case of thermal failure 
\begin{equation}
    p(\tau|f_{0}<t) = \lambda T E_{-\lambda T}(\tau),
\end{equation}
and the total lifetime probability density function
\begin{equation}
    p(\tau) = [1-\exp(-f_{0}\lambda)]\delta(\tau) + \exp(-f_{0}\lambda) \lambda T E_{-\lambda T}(\tau)
\end{equation}
from which we extract the cumulative lifetime distribution via integration over \(\tau\)
\begin{equation}
    P(\tau)= 1 - \lambda T \exp(-f_{0}\lambda) E_{1-\lambda T}(\tau).
    \label{single-fiber-exponential}
\end{equation}  

\nocite{*}

\bibliography{apssamp.bib}

\end{document}